%% file: AI-Paris-article.tex
\documentclass[12pt]{article}
\usepackage{epsfig}
\usepackage{amsmath}
\usepackage{cite}
\input econfmacros.tex
\def\Title#1{\begin{center} {\Large {\bf #1} } \end{center}}

\begin{document}

\Title{Gribov copies and confinement\footnote{Talk held at the Ninth Workshop on Non-Perturbative Quantum Chromodynamics, Paris, June 2007. To appear in the proceedings.}}

\bigskip\bigskip


\begin{raggedright}  

{\it Anton Ilderton\index{Ilderton, A.}\\
School of Mathematics and Statistics\\
University of Plymouth\\
Drake Circus \\
Plymouth PL4 8AA, UK}\\
\bigskip
\texttt{a.b.ilderton@plymouth.ac.uk}
\bigskip\bigskip
\end{raggedright}

\section{Introduction}
We review the construction of locally gauge invariant charges, noting that any such (complete) construction of a quark as an asymptotic state would be in conflict with the observation of only colour-singlets in nature. We show firstly that it must be a non-perturbative effect which prevents such a construction and secondly that is it is closely and explicitly related to the Gribov ambiguity.

Despite a volume of literature on Gribov copies there are few explicit examples. Here we present a new class of well behaved, SU(2) valued, spherically symmetric copies, the non-perturbative nature of which is manifest. This material is based on \cite{Ilderton:2007qy} and references therein.

\section{Coloured charges}
Recall that the Lagrangian fermions $\psi$ of both QED and QCD cannot be identified with observed particles, simply because they are not locally gauge invariant, $\psi\to \psi^U\equiv U^{-1}\psi$. We construct gauge invariant charges $\Psi$ by `dressing' the fermions with a function $h^{-1}[A]$ of the gauge field,
	\begin{equation}
		  \Psi:=h^{-1}[A]\,\psi,
	\end{equation}
an idea which dates back to Dirac \cite{Dirac:1955uv}. The dressed matter field is gauge invariant provided that the dressing factor transforms as
	\begin{equation}\label{defn}
		  h^{-1}[A]\to h^{-1}[A^U] = h^{-1}[A] U,
	\end{equation}
where the gauge field transforms as $A\to A^U\equiv U^{-1}AU+U^{ -1}\partial U.$ This is not enough to specify the dressing factor completely\footnote{There are additional constraints on and contributions to a full dressing. Here we focus only on the `minimal' part which ensures gauge invariance. See \cite{Bagan:1999jf}, \cite{Bagan:1999jk} for details.}. We now identify a general method of constructing dressings through their relation to gauge fixing choices. Let such a choice be denoted $\chi(A)=0$. Given some field $A$ there exists a transformation $h[A]$ into the gauge $\chi\big(A^{h[A]}\big)=0.$ The same must be true for any gauge transformed $A^U$,
	\begin{equation*}
		\chi\bigg(A^{U\,h[A^U]}\bigg)=0.
	\end{equation*}
Now, provided $\chi$ is a good gauge fixing, uniqueness implies $A^{h[A]} = A^{U\, h[A^U]}$ so that
	\begin{equation} 
		h[A]=U h[A^U] \implies h^{-1}[A^U] = h^{-1}[A] U.
	\end{equation}
This is precisely the property (\ref{defn}) we required of our dressings, which may therefore be viewed as the transformations which take any field to a particular gauge slice. Different gauges lead to different dressings and physical interpretations. As an example, it is straightforward to check that the Coulomb gauge in U(1) leads to the dressed fermion
\begin{equation}
	\Psi_e :=\exp \bigg[ig\,\frac{\partial_i A_i}{\nabla^2}\bigg]\,\psi.
\end{equation}
This is Dirac's description of a static electron \cite{Dirac:1955uv}. It is locally gauge invariant, generating a Coulomb field,
\begin{equation}
	E_j\, {\Psi}_e\ket{0} = -\frac{g}{4\pi}\frac{x_j}{r^2}\,{\Psi}_e\ket{0},
\end{equation}
and couples to the photon with the usual strength \cite{Bagan:1999jf, Bagan:1999jk}. Dressing also vastly improves the infra-red behaviour of the field. S-matrix elements of the dressed fermions, for example, are free of soft divergences to all orders in perturbation theory. We would like to reproduce these successes in non-abelian gauge theory, so we look now to the non-abelian Coulomb dressing.  We seek the non-abelian gauge transformation $h[A]$ such that
\begin{equation}
	\partial_i A^h_i \equiv \partial_i\, \bigg( h^{-1} A_i h+\frac{1}{g}h^{-1}\partial_i h\bigg)=0.
\end{equation}
This equation is not as easy to solve as its abelian counterpart, though the solution may be calculated perturbatively order by order in the coupling $g$ \cite{Ilderton:2007qy}. To lowest order we find
\begin{equation}
	\Psi=h^{-1}[A]\psi=\exp\bigg[ g\frac{\partial_i A_i^c}{\nabla^2}T^c\bigg]\psi.
\end{equation}
This term, and all higher terms, vanish when $\partial_i A_i=0$, i.e. $h^{-1}[A]=1$ if $A$ is in Coulomb gauge. This is an important point to which we will later return. What are the properties of $\Psi:=h^{-1}[A]\psi$? It is locally gauge invariant by construction, and calculating the potential between two such objects we find
	\begin{equation}
		\bra{\bar\Psi(y)\ \Psi(x)}\hat{H}\ket{\bar\Psi(y)\ \Psi(x)} =  -\frac{g^2 C_F}{4\pi |x-y|}+\ldots,
	\end{equation}
i.e. we have found the inter-quark potential. Again, higher order terms may be calculated, allowing us to probe screening and anti-screening effects \cite{Bagan:2005qg}. Like the static electron, this dressed object has improved infra-red properties, for example the one loop propagator is gauge invariant and infra-red finite. These good properties lead us to conclude that $\Psi$ is a static quark, and it seems as if we may build gauge invariant coloured charges, at least perturbatively.

Isolated quarks are not observed in nature, and we might expect that there exists a necessarily non-perturbative obstacle to the construction of coloured objects. In the next section we will look for this obstacle.

\section{Gribov copies}
\begin{figure}[htb]
\begin{center}
	\epsfig{file=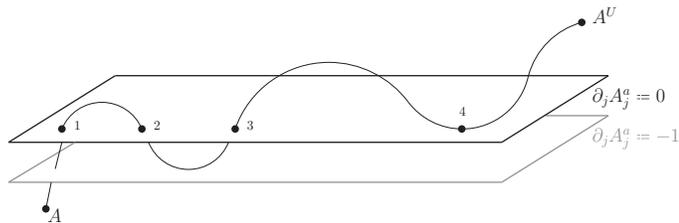,height=1.2in}
\caption{Many gauges multiply intersect (points 1--4) gauge orbits.}
\label{fig:cop}
\end{center}
\end{figure}
Many gauge choices $\chi(A)=0$ fail to select a unique representative field from each gauge orbit, Figure \ref{fig:cop}. The degenerate fields, gauge equivalent yet all lying in the chosen gauge slice, are called Gribov copies \cite{Gribov:1977wm}, \cite{Singer:1978dk}. We are looking for a non-perturbative effect. Recall it is believed that copies are not an issue in perturbation theory, essentially due to appearances of inverse powers of the coupling in their defining equations. There are, however, very few explicit examples of copies in the literature \cite{Henyey:1978qd}, \cite{vanBaal:1991zw}.

We find copies by beginning with a field $A$ in Coulomb gauge, performing a gauge transformation $A\to A^U$, and asking that the new gauge field is also in Coulomb gauge, $\partial_i A^{U}_i=0$ -- if we had a perfect gauge fixing, the only solution would be ${U=1}$. There  are in fact many solutions, and here we present a new class of spherically symmetric SU(2) valued copies and briefly investigate their properties. We refer the reader to \cite{Ilderton:2007qy} for more details on the construction. 

We begin with the observations that gauge fields of the form
\begin{equation*}
	A_i^c(x) = \frac{{  a(r)}-1}{r}\epsilon_{icb}\frac{x^b}{r},
\end{equation*}
for some $a(r)$, are automatically in Coulomb gauge and are spherically symmetric. This symmetry is preserved under SU(2) gauge transformations of the form
\begin{equation*}
  U(x) = \cos\big( gu(r)\big) - i\sin\big( gu(r)\big)\,\frac{\sigma^c x^c}{r}.
\end{equation*}	
We have two degrees of freedom, $u(r)$ and $a(r)$. Requiring $\partial_i A^U_i=0$ implies a differential equation relating $u(r)$ and $a(r)$.
This equation may be written
\begin{equation*}
	a(r) = \frac{r^2u''(r)+2ru'(r)}{\sin \big(2gu(r)\big)}+1-\frac{1}{g}.
\end{equation*}
Rather than treating this as a differential equation for $u(r)$ (which means solving a complicated non-linear differential equation) our  strategy is to choose $u(r)$ and use the equation as an identity for $a(r)$. This gives us a gauge field which automatically obeys both $\partial_iA_i=0$ and $\partial_i A^U_i=0$. We consider only those gauge transformations which approach an element of the centre as $r\to 0$ and $r\to\infty$ \cite{Ilderton:2007qy} and ask that the gauge fields give finite energy configurations. This imposes only the following mild restrictions on $u(r)$:
\begin{eqnarray*}
		u(r)&\sim& \frac{n\pi}{g} + {  c r}\ \text{as}\ r\to0, \\
		u(r)&\sim& \frac{m\pi}{g} + {  \frac{k}{r^2}}\ \text{as}\ r\to\infty,
	\end{eqnarray*}
where $c$ and $k$ are arbitrary constants. There remains an infinite number of allowable $u(r)$'s, giving both small and large transformations. For example, choosing $u(r)=r(1+r^3)^{-1}$, which gives a small gauge transformation, we find the gauge field
\begin{equation*}
	a(r) = \frac{2 r (-7 r^3+r^6+1)}{(1+r^3)^3\sin\big(\frac{2{  g}r}{1+r^3}\big)}+1-{  \frac{1}{g}}.
\end{equation*}
Note the dependence on the coupling $g$ -- this field clearly lies outside of perturbation theory. This is in fact a general feature of our solutions, an appealingly concrete realisation of the statement that the Gribov ambiguity is a non-perturbative phenomenon.

We have constructed a huge class of fields and copies with finite energy (and $L^2$ norm, although this unphysical condition may be relaxed, enlarging the set of copies). Although the copies are manifestly non-perturbative, they may be generated infinitesimally and we may write down copies which are arbitrarily close together. It is therefore clear that these copies lie outside the fundamental domain.

\section{Confinement}
\begin{figure}[htb]
\begin{center}
	\epsfig{file=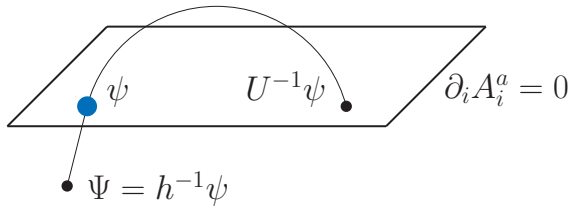,height=1.2in}
\caption{The gauge (in-)dependence of dressed matter.}
\label{fig:conf}
\end{center}
\end{figure}
Having seen that Gribov copies are explicitly a non-perturbative phenomenon we now return to questions of charges and confinement. We saw earlier that physical quarks could be constructed, in perturbation theory, using the Coulomb gauge fixing condition. This construction relied on Coulomb being a `good' gauge fixing condition. Perturbatively this is indeed the case, but non-perturbatively we have Gribov copies which will spoil our construction. We will now see precisely how this occurs.

The static quark dressing depends on $\partial_i A_i$ and is unity when $\partial_i A_i=0$. With reference to Figure \ref{fig:conf} consider the dressed field $\Psi=h^{-1}\psi$. When in Coulomb gauge $h^{-1}=1$ and the dressed fermion $\Psi$ coincides with the Lagrangian fermion $\psi$, as illustrated. Suppose that at this point on the gauge orbit we have a blue quark. By construction our dressed field is gauge invariant and this colour is preserved along gauge orbits-- or would be if not for the Gribov copies. Taking some $A$ we can perform a transformation which takes us to a Gribov copy of $A$, i.e. we can perform a transformation which brings us back into the gauge slice, as illustrated. Under such transformations $\psi\to U^{-1}\psi$ as usual but the dressing does not transform -- it is always unity for a field in Coulomb gauge. We see that $\Psi$ acquires a non-perturbative gauge dependence because of the Gribov copies and thus has no well defined colour. 

We conclude that coloured objects, while they may be constructed in perturbation theory, pick up a gauge dependence non-perturbatively. This arises through the Gribov copies. Such states, therefore, cannot be identified with physical states. The converse statement is that physical states must be `white' -- colour singlets. So, the presence of Gribov copies imply there can be no isolated coloured charges, in other words Gribov copies tell us colour charges are confined.

\section{Conclusions}
The Gribov ambiguity is a degeneracy inherent in many gauge conditions. It has also been shown that to construct a dressed state with well defined colour charge, the boundary conditions on allowed gauge transformations are such that the Gribov ambiguity cannot be avoided \cite{Singer:1978dk}.

The ambiguity is more than just a technical issue to do with over counting of degrees of freedom, having definite physical implications. We have seen that it leads, via the introduction of a non-perturbative gauge dependence to perturbatively invariant states, to the absence of coloured physical states. There remains an open question of how our arguments may be translated into detailed dynamical arguments which will allow us to establish the scale of confinement.

We have added a new class of Gribov copies to the few explicit examples known. A deep question to be addressed is that of the physical significance of the Gribov horizon. To what physical extent, if any, does it matter whether a given configuration lies inside or outside the horizon? We feel the open questions given here are interesting problems worthy of further research.

\def\Discussion{
\setlength{\parskip}{0.3cm}\setlength{\parindent}{0.0cm}
     \bigskip\bigskip      {\Large {\bf Discussion}} \bigskip}
\def\speaker#1{{\bf #1:}\ }
\def\endDiscussion{}


 
\end{document}

%% file: econfmacros.tex
\def\beq{\begin{equation}}
\def\eeq#1{\label{#1}\end{equation}}
\def\eeqn{\end{equation}}

\def\beqa{\begin{eqnarray}}
\def\eeqa#1{\label{#1}\end{eqnarray}}
\def\eeqan{\end{eqnarray}}

\let\bar=\overbar

\def\bra#1{\left\langle{ #1} \right|}
\def\ket#1{\left| {#1} \right\rangle}

\def\Dslash{\not{\hbox{\kern-4pt $D$}}}
\def\dslash{\not{\hbox{\kern-2pt $\del$}}}

\def\msb{{\bar{\ssstyle M \kern -1pt S}}}




%% file: AI-Paris-article.bbl
\begin{thebibliography}{99}

\bibitem{Ilderton:2007qy}
  A.~Ilderton, M.~Lavelle and D.~McMullan,
  JHEP {\bf 0703} (2007) 044
  [arXiv:hep-th/0701168].

\bibitem{Dirac:1955uv}
  P.~A.~M.~Dirac,
  Can.\ J.\ Phys.\  {\bf 33} (1955) 650.

\bibitem{Bagan:1999jf}
  E.~Bagan, M.~Lavelle and D.~McMullan,
  Ann.\ Phys.\  {\bf 282} (2000) 471
  [arXiv:hep-ph/9909257].

\bibitem{Bagan:1999jk}
  E.~Bagan, M.~Lavelle and D.~McMullan,
  Ann.\ Phys.\  {\bf 282} (2000) 503
  [arXiv:hep-ph/9909262].

\bibitem{Bagan:2005qg}
  E.~Bagan, M.~Lavelle and D.~McMullan,
  Phys.\ Lett.\  B {\bf 632} (2006) 652
  [arXiv:hep-th/0510077].

\bibitem{Gribov:1977wm}
  V.~N.~Gribov,
  Nucl.\ Phys.\  B {\bf 139} (1978) 1.

\bibitem{Singer:1978dk}
  I.~M.~Singer,
  Commun.\ Math.\ Phys.\  {\bf 60} (1978) 7.
  
\bibitem{Henyey:1978qd}
  F.~S.~Henyey,
  Phys.\ Rev.\  D {\bf 20} (1979) 1460.

\bibitem{vanBaal:1991zw}
  P.~van Baal,
  Nucl.\ Phys.\  B {\bf 369} (1992) 259.
\end{thebibliography}
